# ENHANCING FILM GRAIN CODING IN VVC: IMPROVING ENCODING QUALITY AND EFFICIENCY

Vignesh V Menon, Adam Wieckowski, Christian Stoffers, Jens Brandenburg, Christian Lehmann, Benjamin Bross, Thomas Schierl, Detlev Marpe

Video Communication and Applications Department, Fraunhofer HHI, Berlin, Germany

**ABSTRACT**

This paper presents an in-depth analysis of film grain handling in open-source implementations of the Versatile Video Coding (VVC) standard. We focus on two key components: the Film Grain Analysis (FGA) module implemented in VVenC and the Film Grain Synthesis (FGS) module implemented in VVdeC. We describe the methodologies used to implement these modules and discuss the generation of Supplementary Enhancement Information (SEI) parameters to signal film grain characteristics in the encoded video sequences. Additionally, we conduct subjective and objective evaluations across Full HD videos to assess the effectiveness of film grain handling. Our results demonstrate the capability of the FGA and FGS techniques to accurately analyze and synthesize film grain, thereby improving the visual quality of encoded video content. Overall, our study contributes to advancing the understanding and implementation of film grain handling techniques in VVC open-source implementations, with implications for enhancing the viewing experience in multimedia applications.

**INTRODUCTION**

In recent years, the proliferation of high-definition video content across various platforms has led to an increased focus on optimizing video coding standards to achieve higher compression efficiency and improved visual quality. The Versatile Video Coding (VVC) standard, developed by the Joint Video Experts Team (JVET) of the International Telecommunication Union (ITU) and the Moving Picture Experts Group (MPEG), represents the latest advancement in video compression technology [1]. VVC offers significant improvements over its predecessors, such as High Efficiency Video Coding (HEVC) [2], regarding compression efficiency, flexibility, and support for emerging multimedia applications [3].

A critical aspect of video coding is handling film grain, which refers to the random noise inherent in film-based video content. Film grain plays a crucial role in the visual aesthetic of many movies and television shows, contributing to the overall texture and atmosphere of the scenes. However, accurately preserving and reproducing film grain during encoding poses several challenges for video coding standards. These challenges stem from film grain's random and temporally uncorrelated nature, its varying characteristics across different films and scenes, and the need to balance preservation with compression efficiency [5].

The handling of film grain in VVC open-source implementations has emerged as a topic of interest among researchers and practitioners in video coding [6, 7]. The Film Grain Analysis (FGA) module implemented in the Fraunhofer Versatile Video Encoder (VVenC) [22] and the Film Grain Synthesis (FGS) module implemented in the Fraunhofer Versatile Video Decoder (VVdeC) [32] are vital components responsible for analyzing and synthesizing film grain in encoded video sequences. Understanding the methodologies and techniques employed in these modules is essential for optimizing film grain handling in VVC implementations and enhancing the visual quality of encoded video content.

This paper comprehensively analyzes film grain handling in VVC open-source implementations, focusing specifically on the FGA module in VVenC and the FGS module in VVdeC and VTM. We describe the methodologies used to implement these modules and discuss the generation of Supplementary Enhancement Information (SEI) parameters to signal film grain characteristics (FGC) in encoded video sequences. Additionally, we conduct subjective and objective evaluations to assess the effectiveness of film grain handling.

**BACKGROUND**

**Overview of VVC**

At the core of VVC is the aim to achieve higher compression efficiency while maintaining or enhancing visual quality [8]. This is accomplished through various innovative techniques and tools that optimize the encoding process and maximize coding efficiency. One critical advancement in VVC is using more sophisticated prediction and transform methods, including enhanced intra-prediction modes, adaptive transform block sizes, improved motion compensation techniques [9, 10, 11], and enhanced in-loop filters [9]. These advancements allow VVC to accurately represent complex video content, resulting in higher coding efficiency and improved compression performance. Furthermore, VVC incorporates several enhancements to address the challenges posed by emerging multimedia applications, such as virtual reality (VR), augmented reality (AR), and immersive video experiences [12, 13]. These enhancements include support for 360-degree video coding, multi-depth video coding, and advanced coding tools for dynamic range and color gamut expansion [14]. By providing robust support for these emerging applications, VVC ensures compatibility with future multimedia technologies and facilitates the delivery of immersive and interactive content experiences.

In addition to its compression efficiency and versatility, VVC also prioritizes scalability and interoperability. The standard accommodates various deployment scenarios, from traditional broadcast and streaming applications to emerging technologies like 5G networks and cloud-based video services [15]. VVC achieves this by offering a scalable coding structure, flexible bitstream syntax, and comprehensive metadata and signaling capabilities support [1].

**Overview of VVenC and VVdeC**

VVenC is an optimized implementation of the VVC standard. Developed by Fraunhofer HHI, it offers high compression efficiency while reducing computational demands. VVenC includes optimizations like SIMD enhancements, improved search algorithms, and multi-threading, enabling faster performance than the VVC test model (VTM). It provides presets to balance encoding speed and quality, achieving significant bitrate reductions compared to the HEVC test model HM [16]. VVenC supports features like frame-level rate control and adaptive quantization and handles both SDR and HDR content [17].

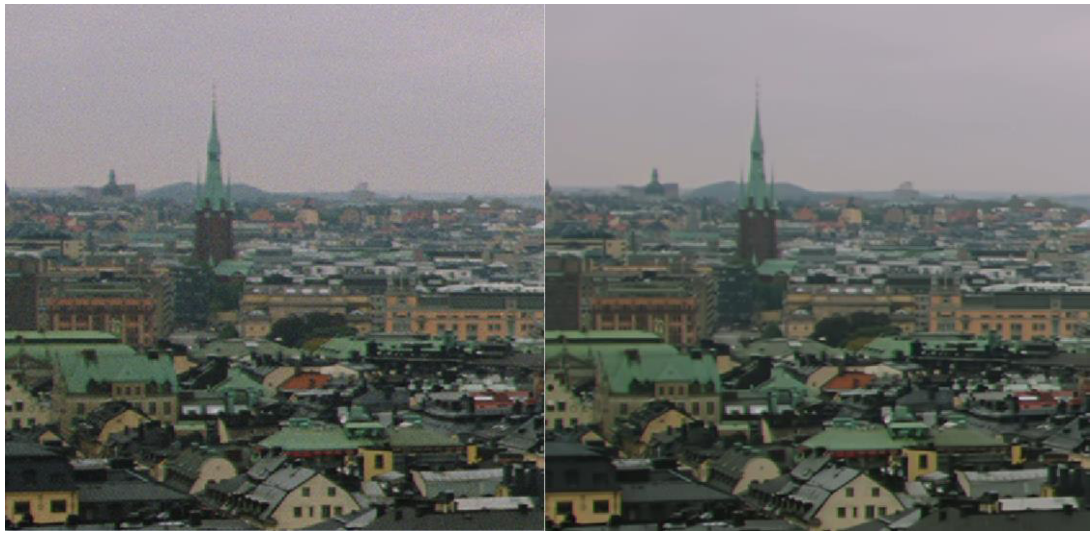

(a) Original video  (b) Encoded at 3 Mbps (*faster* preset)

Figure 1 – Illustration of grain loss in *OldTownCross* sequence when encoded with VVenC.

VVdeC is a software decoding implementation for the VVC standard. VVdeC aims to achieve live decoding efficiently on general-purpose CPUs. It uses advanced techniques like Single Instruction Multiple Data (SIMD) optimizations and multi-threading to manage VVC's added complexity. VVdeC can decode 10-bit HD video at 60 frames per second on modern consumer hardware. VVdeC continues to be optimized for compliance with the final VVC standard while ensuring robust performance.

**Importance of Film Grain Handling**

Film grain is a characteristic feature of film-based video content, originating from the chemical and physical processes involved in capturing images on celluloid film stock. While film grain adds a distinctive texture and aesthetic appeal to movies and television shows, its preservation and faithful reproduction during digital encoding pose significant challenges for video coding standards [18].

One of the primary reasons for the importance of film grain handling in video coding is its impact on visual quality. Film grain contributes to images' perceived sharpness, depth, and texture, enhancing the overall viewing experience for audiences [19, 20], as shown in Figure 1. Moreover, film grain plays a crucial role in cinematographic storytelling and artistic expression. Directors and cinematographers often utilize film grain intentionally to evoke specific emotions, create mood, and establish visual style in their productions. The presence or absence of film grain can significantly impact the tone and atmosphere of a scene, influencing audience perceptions and interpretations of the content.

In digital video coding standards like VVC, effective handling of film grain is essential for ensuring compatibility with legacy content and seamless integration with existing workflows [21]. By accurately analyzing and synthesizing film grain in encoded video sequences, VVC enables broadcasters, content creators, and distributors to maintain consistency and fidelity across different generations of video content [22, 23]. Additionally, robust film grain handling capabilities enhance the versatility and applicability of VVC in diverse multimedia applications, including film restoration, archival digitization, and high-quality content delivery.

## FILM GRAIN ANALYSIS IN VVENC

As shown in Figure 2, the process begins with extracting film grain from the source video using sophisticated denoising techniques. VVenC [24] uses the motion-compensated temporal filter (MCTF), which effectively separates film grain from other video components while preserving essential motion information [25]. The MCTF leverages motion estimation and compensation to identify and attenuate the temporal variation of film grain, resulting in a denoised video signal with enhanced clarity [26].

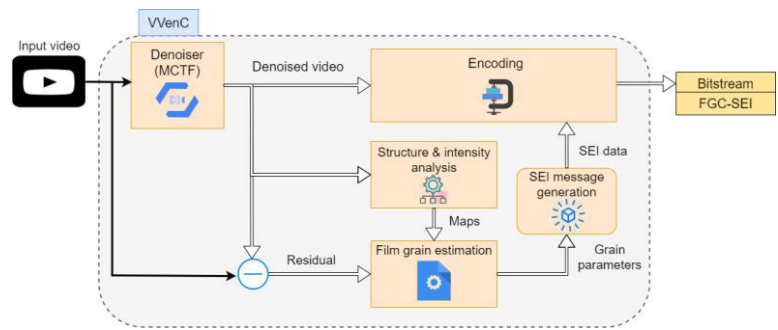

Figure 2 – FGA workflow implemented in VVenC.

Following the extraction of film grain, the next step is quantifying its properties. This involves analyzing parameters such as grain size, intensity, distribution, and temporal variation. Initially, the algorithm identifies and masks flat and low-complexity regions in the video frame using edge detection and morphological operations. The identified areas are then analyzed to estimate the film grain by subtracting the filtered frame from the original, capturing the residual grain. This residual is transformed using a Discrete Cosine Transform (DCT), and the resulting coefficients are used to evaluate the variance of the grain pattern. The algorithm fits a polynomial function to the variance data, smoothing it and performing Lloyd-Max quantization to derive scaling factors. These factors are then refined and quantized to match the bit depth requirements. Finally, as supported by the SEI message format, the algorithm defines intensity intervals and scaling factors, ensuring they are correctly set and scaled down to 8-bit ranges. In the future, statistical analysis and machine learning-based approaches may be utilized to characterize film grain properties accurately, as discussed in [27, 28].

Once the film grain characteristics are determined, they are integrated into the encoding process through the FGC SEI messages. These messages, specified in VSEI standard for VVC, transmit the model parameters to the decoder, enabling it to replicate the film grain during decoding [29, 30, 31]. The methodology ensures the accurate transmission of film grain parameters through efficient encoding and decoding strategies, minimizing overhead while preserving fidelity.

## SUPPLEMENTARY ENHANCEMENT INFORMATION PARAMETERS FOR FILM GRAIN

It is important to note that VVC's FGC SEI specification only provides the syntax to transmit model parameters to the decoder without detailing methods for parameter estimation or FGS [30, 31]. This allows the encoder to use the FGC SEI message to characterize film grain present in the source video material, which might be removed by pre-processing filtering or lossy compression, or to synthesize film grain on decoded images even when it is not present in the original content [29]. Specifically, tuned parameters can mask compression artifacts, even if no film grain exists in the original video. The FGC SEI message facilitates the signaling of various film grain simulation models, blending modes, bit depths, transfer characteristics, chroma processing options, and model-specific parameters. These model-specific parameters for SMPTE-RDD5 [32], which are part of the FGC SEI message, are detailed in Table 1. In the following, we provide a brief overview of the semantics of these syntax elements as defined in VSEI and following the SMPTE-RDD5 model [32].

The FGC message accommodates two synthesis approaches: autoregressive modeling of film grain and frequency filtering modeling. When the *film_grain_model_id* is set to 0, it signifies the use of the frequency filtering model. The presence of the *separate_colour_description_present_flag* indicates that the film grain color space matches the encoding color space. Additive blending [33] is employed when the *blending_mode_id* is set to 0, which defines the blending mode used to blend the simulated film grain with the decoded images. The *log2_scale_factor* is utilized to scale the film grain scaling parameter provided with *comp_model_value[c][i][0]*, effectively restricting the range of grain strength and encouraging the use of meaningful values. Subsequently, the *comp_model_present_flag* is specified for each color component (e.g., luma and two chroma components). A *comp_model_present_flag[c]* equal to 1 denotes the presence of syntax elements for modeling film grain on color component c within the SEI message, prompting the application of FGS exclusively to components with this flag set to 1. Conversely, if *comp_model_present_flag[c]* is 0, the respective component's FGS and blending are skipped.

For each color component for which the synthesis process is invoked, it defines the number of intensity intervals (*num_intensity_intervals_minus1[c]*) and the number of model values (*num_model_values_minus1[c]*). Since film grain depends on an image's local characteristics, different scales of the film grain can be applied for different intensities or intensity intervals of an input image to get the appropriate intensity of the film grain before finally blending it into the input image. After creating a film grain pattern following the frequency filtering model, scaling to the proper intensity based on the scaling factor (SF) is performed. It determines the level at which the film grain will be perceived in the final image, and by doing that, we ensure that the film grain is simulated at the correct scale. Currently, SMPTE-RDD5 and FGC SEI message, as defined in VVC, includes the following parameters that define the film grain scaling function, which is applied at the decoder/synthesis side:

1. *intensity_interval_upper_bound[c][i]* less than *intensity_interval_lower_bound[c][i+1]*
2. *comp_model_value[c][i][0]*,

| Syntax element | Value |
|---|---|
| film_grain_model_id | 0 |
| separate_colour_description_present_flag | 0 |
| blending_mode_id | 0 |
| log2_scale_factor | 2-7 |
| for (c=0; c< 3; c++ ) | |
|   comp_mode_present_flag[c] | 0-1 |
| for (c=0; c< 3; c++ ) { | |
|   If (comp_model_present_flag[c]){ | |
|     num_intensity_intervals_minus1[c] | 0-9 |
|     num_model_values_minus1[c] | 0-2 |
|     for(i=0; i < num_intensity_intervals_minus1[c]+1; i++){ | |
|       Intensity_interval_lower_bound[c][i] | $0-2^b-1$ |
|       Intensity_interval_upper_bound[c][i] | $0-2^b-1$ |
|       for(j=0; j < num_intensity_intervals_minus1[c]+1; j++) | |
|         comp_model_value[c][i][j] | |
|     } | |
|   } | |
| } | |
| film_grain_characteristics_persistence_flag | 0 |

Table 1 – FGC SEI syntax parameters and semantics for SMPTE RDD5 [30].

where comp model value[c][i][0] represents the scaling factor SF, separately defined for each color component c and intensity interval i. The given representation leads to the piecewise constant scaling function (a.k.a, stepwise scaling function). Given the SMPTE-RDD5 specification, intensity intervals cannot overlap (multi-generational film grain is not allowed in SMPTE-RDD5). However, the general FGC SEI message design, as in VSEI, does not prevent using multi-generational film grain.

The minimum number of model parameters that need to be transferred is 1 (see *num_model_values_minus1[c]*, which specifies the number of model values present for each intensity interval). In such a way, for each intensity interval i, it is necessary to define at least SF (*comp_model_value[c][i][0]*). It corresponds to the standard deviation of the Gaussian noise used in the film grain pattern generation process, and it ranges within the interval [0; $2^b$-1]. Typically, for the frequency filtering model defined by SMPTE-RDD5, two other parameters are required, namely horizontal high cut-off frequency (*comp_model_value[c][i][1]*) and vertical high cut-off frequency (*comp_model_value[c][i][2]*). They represent cut-off frequencies of a low-pass filter used to generate film grain patterns. According to SMPTE-RDD5, they should be in the range [2; 14]. In some cases, e.g., when *num_model_values_minus1[c]* is 0, the cut-off frequencies are not encoded in the bitstream and are inferred to be eight at the decoder side. If, however, only one cut-off frequency is encoded in (*num_model_values_minus1[c]* is 1), the other one is implicitly derived, and two are considered the same. It is to be noted that SMPTE-RDD5 defines low-pass filtering, and it represents just one implementation variant of the frequency filtering approach. However, another type of filtering can be used in general. FGC SEI allows up to six model parameters to be defined (*num_model_values_minus1[c]* can be up to 5 for general SEI design as described in VSEI). In such a way, for example, one can use band-pass filtering. By SMPTE-RDD5, the FGC SEI message is inserted at each frame, which is indicated by setting the *film_grain_characteristics_persistence_flag* to 0. It also means the FGC SEI message is applied only to the current decoded frame. Inserting the SEI message on each frame has advantages to ensuring a safe and bit-accurate process in tricky mode, e.g., fast-forward display mode.

**FILM GRAIN SYNTHESIS IN VVDEC**

The work in SMPTE-RDD5 [32] provides an in-depth look at FGS, which is part of the decoder side of the video distribution chain, implemented in VVdeC [34]. Although this process is defined for the H.264 standard [36], it is compatible with VVC without modifications since both support the same metadata [29]. Only minor adjustments are needed to support bit depths higher than 8-bit. This method offers a more precise FGS specification than VSEI [30]. It is based on filtering in the frequency/transform domain, which involves filtering random noise to simulate the film grain pattern. In this model, film grain patterns are generated in the frequency domain using a pair of cut-off frequencies that define a low-pass filter. These patterns are then scaled to the proper intensity before blending into the image.

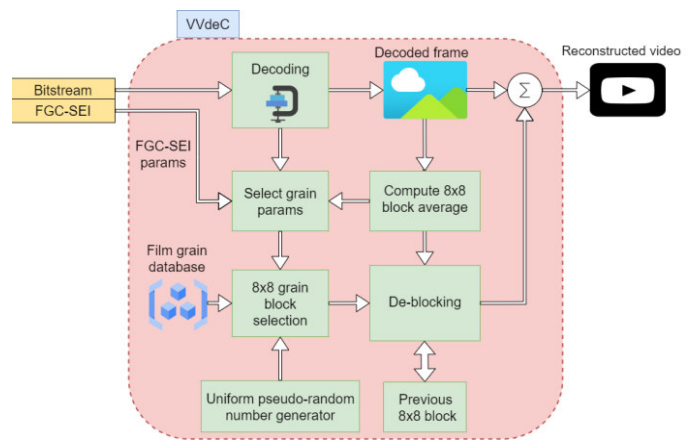

Figure 3 – FGS and blending for SMPTE-RDD5 [32] implemented in VVdeC.

The workflow for film grain simulation and blending in SMPTE-RDD5 is illustrated in Figure 3. It is described in [7] but included in this paper to make it self-contained. This process involves specifying a database of film grain patterns, a uniform pseudo-random number generator, and a precise sequence of operations, such as deblocking and selecting film grain parameters, all defined within the specification. In the proposed implementation, the film grain database is created offline for all combinations of cut-off frequencies. SMPTE-RDD5 defines a pre-computed set of transformed pseudo-random numbers stored for future use. These pre-computed sets are obtained using an integer approximation of the floating-point DCT, as specified in SMPTE-RDD5 [32].

The next step is to perform low-pass filtering on the 64 x 64 block of transformed pseudo-random values, denoted as *B*. Each film grain pattern is synthesized using different pairs of cut-off frequencies. The horizontal and vertical high cut-off frequencies ($h_C$ and $v_C$) define the film grain pattern. Low-pass filtering is done by zeroing all block B coefficients such that $x > h_C$ or $y > v_C$, where x and y are horizontal and vertical coordinates within the block. After filtering, the inverse DCT is performed to obtain the block *b'*, representing the film grain pattern.

Different film grain patterns for various cut-off pairs can be pre-computed, generating a comprehensive database of film grain blocks/patterns. The database includes all possible combinations of cut-off frequencies, resulting in *h* x *v* x 64 x 64 film grain samples. The $h_C$, $v_C$, and values for *h* and *v* for the 64x64 block are created as follows:

$$h_C = ((h+3) <<2) -1; h = comp\ model\ value[c][i][1]-2$$

$$v_C = ((v+3) <<2) -1; v = comp\ model\ value[c][i][1]-2$$

Here, << represents the left bit-shift operation.

Once the film grain database is created and an FGC SEI message is received, film grain simulation can begin. To choose a particular pattern from the film grain database, the 8x8 block average is calculated to determine the interval to which the average value of the currently processed 8x8 block belongs. This block is part of the processed image, usually a decoded frame. The average value is compared with the SEI message's intensity interval parameters to identify the intensity interval. Based on this average value and the received FGC SEI parameters, FGC parameters are selected, including the scaling parameter and cut-off frequencies. The chosen cut-off frequencies are used to access the film grain database.

Film grain is added to the image on an 8 x 8 block basis. Random 8x8 blocks are selected from the 64x64 FG block created earlier, using a pseudo-random number generator to define an offset from the origin of the 64x64 FG block, ensuring bit-exact simulation. Adding film grain on an 8x8 basis, rather than using the entire FG block directly, provides the randomness of the film grain, avoiding repetitive patterns that could degrade visual quality. The appropriate intensity of the film grain is achieved through additional scaling as follows:

$$b' = comp\_model\_value[c][i][0] \times b'$$

$$fg\_final = b' >> (log2\_scale\_factor + 6)$$

$$output\_frame = Clip(\ 0,\ (1 << bitdepth) - 1,\ decoded\ frame + fg\_final\ ),$$

Here, *log2_scale_factor* is used to scale the film grain parameter, and an additional downscaling by six is performed since the pre-computed transformed pseudo-random values are scaled up. The >> operator represents the right bit-shift, << is the left bit-shift, and Clip limits the range of output pixel values after blending the final film grain pattern with the decoded frame. Deblocking is also performed to smooth the edges. The input image can be processed block by block, using 8x8 granularity, in raster scan order or any other convenient manner.

# EXPERIMENTAL VALIDATION

## Experimental setup

We conducted our experiments on an AMD EPYC 7502P processor with 32 cores, running each instance of VVenC v1.12 [24] using eight CPU threads and enabling adaptive quantization, using presets *faster* and *medium*. For evaluation, all video sequences were downscaled to 1920x1080 resolution at 8-bitdepth. Videos were encoded using two-pass rate control [34] at bitrates {1.0, 1.5, 2.4, 3.4, 4.5, 5.8, 7.8, 9.0} Mbps. FGS implemented in VTM and VVdeC (using four CPU threads) [35] decoders are compared.

## Objective evaluation

Since the film grain is filtered out before encoding, the proposed toolchain lowers the bitrate needed to achieve a similar perceptual quality. However, traditional metrics such as PSNR and MS-SSIM are not suitable for evaluating the perceptual quality of film grain coding due to their lack of texture sensitivity. PSNR and MS-SSIM are particularly sensitive to noise, penalizing the addition of film grain and leading to lower scores despite improvements in perceptual quality, as observed in Table 2. VMAF [37], while more advanced, is not specifically trained to evaluate the perceptual quality of VVC-coded videos [38]. This limitation reduces its effectiveness in accurately assessing the enhancements brought by film grain coding.

Given these constraints, specialized metrics focusing on texture enhancement, perception of controlled noise, and overall film-like appearance would be more appropriate for evaluating film grain coding. Future work should aim to develop and utilize metrics that incorporate human perception aspects and consider both texture fidelity and noise [40]. Such metrics would better assess the quality enhancements that film grain brings to video content.

| Preset | Video | BD-PSNR [in dB] | BD-M3SSIM [in dB] | BD-VMAF |
|---|---|---|---|---|
| faster | BQTerrace | -1.50 | -3.23 | -3.06 |
| | CrowdRun | -0.28 | -0.68 | -0.92 |
| | InToTree | -0.95 | -1.95 | -2.47 |
| | OldTownCross | -1.69 | -3.64 | -3.53 |
| | ParkScene | -1.01 | -1.86 | -1.09 |
| medium | BQTerrace | -1.67 | -3.67 | -3.43 |
| | CrowdRun | -0.33 | -0.96 | -1.24 |
| | InToTree | -1.05 | -2.07 | -2.75 |
| | OldTownCross | -1.78 | -3.37 | -2.58 |
| | ParkScene | -1.22 | -2.29 | -1.32 |

Table 2 – Bjontegaard-delta [39] quality results with FGA for the *medium and faster* presets using VVenC.

## Subjective evaluation

In scenarios with limited data capacity, such as low bitrate encoding, achieving optimal visual quality poses significant challenges. Compression artifacts, such as blocking and banding, become more pronounced under these constraints, as illustrated in Figure 4. These subfigures compare different encoding scenarios: the original video (top-left), the video encoded at 0.8 Mbps (top-right), and the video encoded at 0.8 Mbps with FGS enabled (bottom). At low bitrates, noticeable compression artifacts, such as blocking and banding, begin to degrade the visual quality of the video. These artifacts are due to the aggressive compression required to fit the video within the limited bitrate, leading to a loss of fine details and the introduction of visual distortions. FGS is employed to mitigate these issues. FGS introduces controlled noise that mimics the natural film grain characteristics of the original

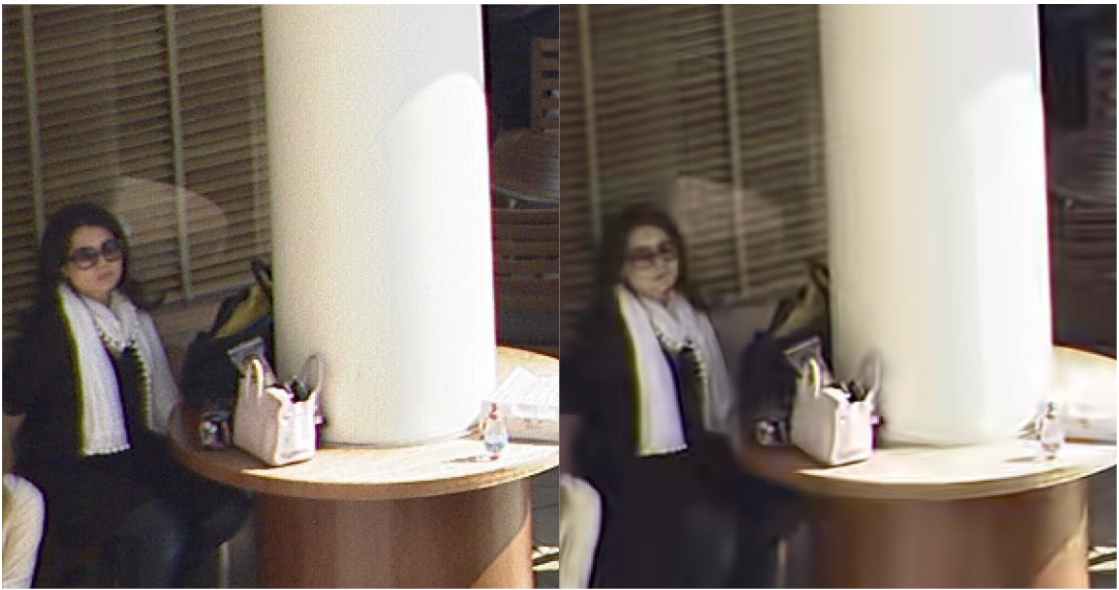

(a) Original video        (b) Default encoding (0.8Mbps)

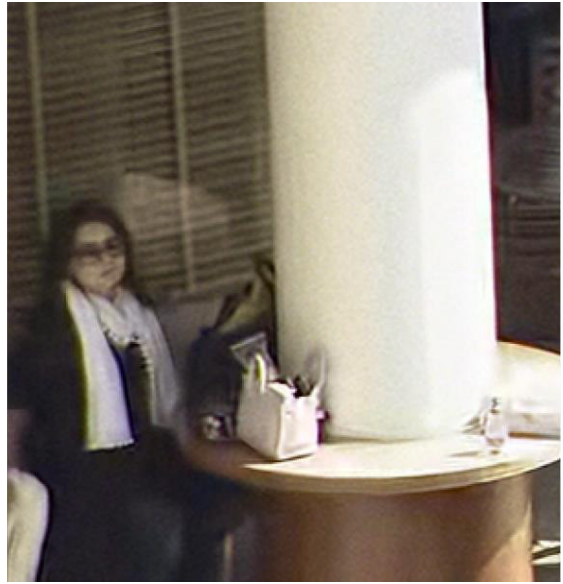

(c) Encoding with FGA and FGS (0.8 Mbps)

Figure 4 – A cropped frame of *BQTerrace* encoded using *faster* preset. The compression artifacts are observed to be masked, especially on the window blinds.

video sequence. This synthesized grain acts as a visual distraction, effectively camouflaging compression-related imperfections. A similar observation is made for encoding another test sequence, shown in Figure 5.

**Encoding and decoding time**

Table 3 illustrates the increase in runtime complexity with FGA for the *medium* and *faster* presets using VVenC. Owing to our optimization that FGC parameters are estimated for frames only after a gap of at least one GOP size and FGA being carried out only for the frames where MCTF is enabled by default, the additional computational overhead is negligible.

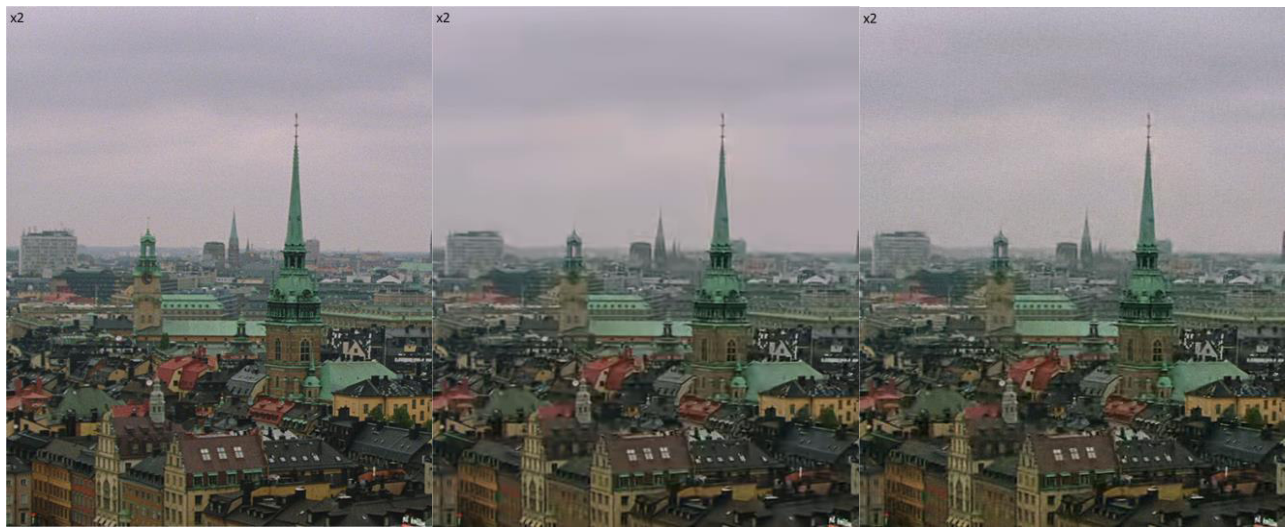

    (a) Original video        (b) Default encoding        (c) Encoding with FGA & FGS

Figure 5 – A cropped frame of *OldTownCross* encoded at 0.6 Mbps using *faster* preset.

Table 4 compares the decoding speeds (in frames per second) of VTM and VVdeC decoders, both with and without FGS. The results show that while FGS implementation is consistent across both decoders, its impact on decoding speed varies significantly. For the VTM decoder, the speed reduction with FGS enabled is minimal, as seen with the "*CrowdRun*" sequence [41], where the speed decreases slightly from 1.88 to 1.77 fps. VVdeC experiences a decrease in speed, dropping from 112.76 fps to 95.07 fps for the same sequence. However, on average, VVdeC (FGS) is approximately 60 times faster than VTM (FGS) and can handle real-time decoding. The optimization of FGS within the VVdeC decoder is a work in progress and remains a focus for future improvements. Efforts are ongoing to enhance VVdeC's efficiency in handling FGS to achieve better speed, ensuring that the perceptual benefits of FGS can be realized without substantial compromises in decoding speed.

| Video | ΔT (medium) | ΔT (faster) |
|---|---|---|
| BQTerrace | -0.76% | -0.35% |
| CrowdRun | -0.69% | -0.95% |
| InToTree | -0.62% | -3.25% |
| OldTownCross | -0.76% | 2.21% |
| ParkScene | -0.78% | 1.27% |

Table 3 – Runtime increase with FGA for the *medium and faster* presets using VVenC.

| Video | VTM | VTM (FGS) | VVdeC | VVdeC (FGS) |
|---|---|---|---|---|
| BQTerrace | 1.86 | 1.75 | 114.03 | 96.17 |
| CrowdRun | 1.88 | 1.77 | 112.76 | 95.07 |
| InToTree | 1.83 | 1.72 | 111.82 | 94.68 |
| OldTownCross | 2.01 | 1.88 | 117.28 | 98.65 |
| ParkScene | 1.58 | 1.49 | 91.72 | 79.60 |

Table 4 – Comparison of decoding speeds (in fps) of VTM and VVdeC decoders.

## CONCLUSIONS

This paper presented the multifaceted impact of FGA and FGS on the encoding, decoding, and subjective perception of video content for VVC-based implementations. In scenarios such as low-bitrate encoding, FGS emerges as a valuable tool in mitigating compression artifacts by introducing controlled noise that mimics natural film grain characteristics, effectively camouflaging compression artifacts. However, integrating FGS introduces computational overhead, as evidenced by increased decoding time in our experiments. Despite this, our objective evaluation emphasizes the potential of FGS to lower the required bitrate while maintaining perceptual quality, underscoring its significance in video encoding workflows. Furthermore, our comparison of decoding speeds between VTM and VVdeC decoders underscores the need for ongoing optimization efforts to enhance the efficiency of VVdeC in handling FGS. Ultimately, our findings underscore the complex interplay between encoding techniques, perceptual quality, and computational efficiency, highlighting avenues for future research and development in video coding and processing.